\documentclass[referee]{raa}            % referee version: for submission

\usepackage{color}
\usepackage{graphicx,times}             %for PS/EPS graphics inclusion, new
\usepackage{natbib}
\usepackage{amssymb,amsmath}
\bibpunct{(}{)}{;}{a}{}{,}
\usepackage[pagebackref=true]{hyperref}
\usepackage{caption}
\begin{document}

  \title{Enhancing Cosmological Constraints by Two-dimensional $\beta$-cosmic-web Weighted Angular Correlation Functions
}

   \volnopage{Vol.0 (20xx) No.0, 000--000}      %%preserved for Editor. DOn't remove!
   \setcounter{page}{1}          %%starting page, preserved for Editor. DOn't remove!

      \author{Fenfen Yin
      \inst{1,2}
   \and Liang Xiao
      \inst{2}
   \and Wenying Du
      \inst{2}
    \and Zhujun Jiang
      \inst{2}
    \and Zhiwei Min
      \inst{2} 
    \and Jaime Forero-Romero
     \inst{3}
    \and Jiacheng Ding
     \inst{4}
     \and Le Zhang
     \inst{2,6}
    \and Xiao-Dong Li
    \inst{2,5,6}
   }
 
 \institute{Department of Physics and Electronic Engineering, Tongren University, Tongren 554300, China\\
        \and
            School of Physics and Astronomy, Sun Yat-Sen University, Zhuhai 519082, China; {\it zhangle7@mail.sysu.edu.cn}; {\it lixiaod25@mail.sysu.edu.cn}\\
        \and
             Departamento de Física, Universidad de los Andes, Cra. 1 No. 18A-10 Edificio Ip, CP 111711, Bogotá, Colombia; {\it je.forero@uniandes.edu.co}\\
        \and
            Shanghai Astronomical Observatory, Chinese Academy of Sciences, No. 80 Nandan Road, Shanghai 200030, China; {\it dingjch@shao.ac.cn} 
         \and 
         Peng Cheng Laboratory, No. 2, Xingke 1st Street, Shenzhen 518000, China
         \and 
         CSST Science Center for the Guangdong–Hong Kong–Macau Greater Bay Area, SYSU, Zhuhai 519082, China\\
\vs\no
   {\small Received 20xx month day; accepted 20xx month day}}

\abstract{In this study, we investigate the potential of mark-weighted angular correlation functions (MACFs), which integrate $\beta$-cosmic-web classification with angular correlation function analysis to improve cosmological constraints. Using SDSS DR12 CMASS-NGC galaxies and mock catalogs with $\Omega_m$ varying from 0.25 to 0.40, we assess the discriminative power of different statistics via the average improvement in chi-squared, $\Delta \overline{\chi^2}$, across six redshift bins. This metric quantifies how effectively each statistic distinguishes between different cosmological models. 
Incorporating cosmic-web weights leads to substantial improvements. Using statistics weighted by the mean neighbor distance ($\bar{D}_{\rm nei}$) increases $\Delta \overline{\chi^2}$ by approximately 40\%-130\%, while applying inverse mean neighbor distance weighting ($1/\bar{D}_{\rm nei}$) yields even larger gains, boosting $\Delta \overline{\chi^2}$ by a factor of 2-3 compared to traditional unweighted angular statistics. These enhancements are consistent with previous 3D clustering results, demonstrating the superior sensitivity of the $\beta$-weighted approaches. Our method, based on thin redshift slices, is particularly suited for slitless surveys (e.g., Euclid, CSST) where redshift uncertainties limit 3D analyses. This study also offers a framework for applying marked statistics to 2D angular clustering.
\keywords{Cosmology: cosmological parameters – large-scale structure of Universe; Methods: statistical
}
}
   \authorrunning{Fenfen Yin et al.}
   %Liang Xiao,  Zhiwei Min, Zhujun Jiang, Wenying Du, Jaime Forero-Romero, Jiacheng Ding,Le Zhang \& Xiao-Dong Li }   

   \titlerunning{Enhancing Cosmological Constraints} %by Two-dimensional $\beta$-cosmic-web Weighted Angular Correlation Functions}   

   \maketitle
%________________________________________________ sections below
%
\section{Introduction}
\label{sect:intro}
The cosmic structure originated from early tiny density perturbations. Under the combined influence of gravity and accelerated expansion driven by dark energy, these perturbations gradually give rise to the observed complex and significantly nonlinear structure, which exhibits a wide variety of morphologies and properties at different scales.
 
The spatial distribution of matter in the universe manifests as a complex, hierarchical network structure termed the ``cosmic web''. This anisotropic pattern emerges from gravitational collapse processes dominated by DM dynamics, and has constituted a major focus of cosmological research since its first identification in the late 20th century~\citep{Bardeen+etal+1986,Lapparent+etal+1986,Huchra+etal+2012,Tegmark+etal+2004,Guzzo+etal+2014,Suárez+etal+2021,Bond+etal+1996}.

The study of the cosmic web relies on large-scale galaxy surveys. Currently, several large-scale survey projects have been successfully implemented, such as the Sloan Digital Sky Survey (SDSS)~\citep{York+etal+2000,Abazajian+etal+2003,Abazajian+etal+2004,Eisenstein+etal+2005,Gott+etal+2005,Adelman+etal+2006,Percival+etal+2007,Adelman-McCarthy+etal+2008,Anderson+etal+2012,Sánchez+etal+2012,Sánchez+etal+2013,Anderson+etal+2014,Ross+etal+2015,Beutler+etal+2017,Sánchez+etal+2017,Alam+etal+2017,Chuang+etal+2017}, the 2-degree Field Galaxy Redshift Survey (2dFGRS)~\citep{Colless+etal+2001}, the 6-degree Field Galaxy Redshift Survey (6dFGRS) ~\citep{Beutler+etal+2012,Beutler+etal+2011} and the WiggleZ Dark Energy Survey~\citep{Parkinson+etal+2012}, providing valuable data support for unveiling the mysteries of the cosmic web.

Various methods have been developed to describe and analyze the cosmic web. The most popular techniques for describing the cosmic web include the density-based classification~\citep{Klypin+etal+1997,Springel+etal+2001,Knollmann+Knebe+2009}, 
T-web~\citep{Hahn+etal+2007, Suarez-Perez+etal+2021,Forero_Romero+etal+2009},  V-web~\citep{Hoffman+etal+2012,Forero_Romero+etal+2014}, DisPerSE~\citep{Sousbie+2011}, skeleton algorithms defined based on Lagrangian fluid dynamics~\citep{Feldbrugge+etal+2018},  and so on. These tools explore the properties of the cosmic web by analyzing the velocity field, density field, or the intuitive geometric distribution of the cosmic web. 

Among the various cosmic web descriptions, the $\beta$-skeleton method~\citep{Fang+etal+2019,Suarez-Perez+etal+2021} utilizes an algorithm originates from the fields of computational geometry and geometric graph theory to describe the properties of the cosmic web. This algorithm constructs a graph to describe the {\it connectivity} properties from a set of $n$-dimensional spatial points, and has been applied in a variety of fields such as machine learning, visual perception, image analysis and pattern recognition~\citep{Edelsbrunner+etal+1983,Amenta+etal+1998,Zhang+King+2002}. \cite{Fang+etal+2019} firstly applied this algorithm to the cosmic large-scale structure (LSS), demonstrating that the cosmic web can be intuitively reconstructed from galaxy spatial distributions, enabling effective visualization of this fundamental structure. \cite{García+etal+2020} explored $\beta$-cosmic-web and found it reveals significant advantages in characterizing sparsely distributed galaxies. Based on the cosmic web, they further proposed a measure of entropy for describing the complexity of large-scale structures, and built up the relationship between the $\beta$-cosmic-web and the traditional T-web via machine learning methods. Recently, \cite{Yin+etal+2024} points out that the environmental information of $\beta$-cosmic-web weighting schemes can be utilized to build up mark weighted correlation functions, which can substantially improve the accuracy of the constraints on the cosmological parameters compared to the traditional two-point correlation function.

In the next decade, stage-IV surveys such as Dark Energy Spectroscopic Instrument (DESI)~\citep{DESI+etal+2016, DESI+etal+2022,Adame+etal+2025}, Large Synoptic Survey Telescope (LSST)~\citep{LSST+etal+2009, Ivezic+etal+2019, Crenshaw+etal+2025,Kumar+etal+2025}, Euclid Space Telescope~\citep{Laureijs+etal+2011, Euclid+Scaramella+etal+2022,Euclid+Mellier+etal+2024,Cuillandre+etal+2024,Euclid+Cropper+etal+2024, Euclid+Jahnke+2024}, and China Space Station Telescope (CSST)~\citep{Zhan+etal+2011,Cao+etal+2018,  Gong+etal+2019, Zhan+etal+2021,Fu+etal+2023, Luo+etal+2024,Sui+etal+2025,Shi+etal+2025} will provide deep, wide-field observations in much wider range of redshift, producing unprecedented datasets for cosmic web studies. While these surveys provide abundant data for studying the cosmic evolution, the analysis of the data faces new  challenges. The stage-IV surveys will unveil complex nonlinear structures, which require more advanced analysis methods. Moreover, some stage-IV surveys (e.g. CSST and Euclid) employ slitless grating spectrometer, which efficiently expand the survey volume/depth but at the cost of reduced redshift measurement precision. Because this trade-off may significantly affect data analysis~\citep{Gong+etal+2019, Xiao+etal+2023, Gu+etal+2024, Euclid+Le+etal+2025}, the methodology must be adjusted to address the resulting challenges of redshift uncertainty.

One possible way to avoid the difficulties brought by redshift uncertainty is to use the two-point angular correlation function (2PACF)~\citep{Connolly+etal+2002,Wang+etal+2013, Carvalho+etal+2016, Alcaniz+etal+2016,Carvalho+etal+2018,Carvalho+etal+2020,Venville+etal+2024,Franco+etal+2024,Euclid+Duret+etal+2025,Wu+etal+2025}, which analyzes cosmic large-scale structures without requiring precise redshift information and is ideal for photometric surveys. Yet the limitation of the 2PACF is that it only captures Gaussian features of the density field, becoming increasingly limited as gravitational collapse enhances non-Gaussianity. Recently, mark-weighted correlation functions (MCFs) have been extensively studied~\citep{White+etal+2016, Satpathy+etal+2019, Xiao+etal+2022, Yang+etal+2020, Storey-Fisher+etal+2024, Massara+etal+2024, Xu+etal+2025}. MCFs extract non-Gaussian information by weighting galaxy clustering statistics with galaxy properties, offering a more complete characterization of nonlinear structure. This simple yet powerful method has proven effective in probing the complex features of LSS.

In this study, we introduce a novel mark-weighted statistical measure, the mark-weighted angular correlation functions (MACFs), which combine $\beta$-cosmic-web classification with angular correlation function analysis. This approach retains sensitivity to non-Gaussian features in the LSS while reducing sensitivity to redshift uncertainties.

The study is organized as follows: In Section 2, we briefly introduce the datasets used in this analysis. In Section 3, we detail introduce the methodology of building up $\beta$-cosmic-web from the 2D galaxy distribution, as well as the methodology for the related MACFs. The results and findings are presented in Section 4. We summarize and conclude in Section 5.

\section{Datasets}

To validate the sensitivity of MACFs with $\beta$-cosmic-web weights to cosmological parameters, we compare measurements from the SDSS BOSS DR12 data with those from mock catalogs. The mock simulations are produced using COmoving Lagrangian Acceleration (COLA) fast algorithm~\citep{Tassev+etal+2013,Wang+Shi+etal+2024}, and then calibrated using the subhalo abundance matching (SHAM) methods~\citep{Ding+etal+2024}, so that to accurately reproduce CMASS galaxy statistics. For assessing the power of statistics, we compute covariance matrices from Multidark PATCHY (PerturbAtion Theory Catalog generator of Halo and galaxY distributions)~\citep{Kitaura+etal+2016} simulations. 

\subsection{The BOSS CMASS Galaxies}
\label{sect:Obs}

The Baryon Oscillation Spectroscopic Survey (BOSS) project is a crucial component of SDSS-III~\citep{Eisenstein+etal+2011, Bolton+etal+2012, Dawson+etal+2012}. Its primary scientific goal is to precisely measure expansion and structure growth history via the BAO measurements from the spatial distribution of luminous red galaxies (LRGs) and quasars~\citep{Eisenstein+etal+2001}. To do this, redshift information for about 1.5 million cosmic galaxies within a sky area of approximately 10,000 square degrees is accurately measured. These galaxies are divided into two samples of LOWZ and CMASS. The LOWZ galaxy sample includes mainly brightest red galaxies at $z\leq$0.4, while the CMASS sample targets a population of galaxies at higher redshifts, many of which are also in the category of luminous red galaxies (LRGs). 

To explore the statistical properties of MACFs, the data must be projected from the three-dimensional space onto a two-dimensional angular space. For simplicity, in this analysis we only use a sub-dataset of the BOSS DR12 CMASS NGC galaxies lying in the redshift range of $z\in [0.45, 0.55]$. The process is done by the following steps:
\begin{itemize}
    \item first, we divide the three-dimensional galaxy sample into six overlapping redshift intervals, each with a thickness of $\Delta z=0.05$. The intervals are $z\in [0.45, 0.50]$, $[0.46, 0.51]$, $[0.47, 0.52]$, $[0.48, 0.53]$, $[0.49, 0.54]$, $[0.50, 0.55]$, respectively.
    \item next, we project the galaxies in each redshift shell onto a two-dimensional plane at the shell's central redshift. The central redshifts for the shells are: $z_{c}=0.475, 0.485, 0.495, 0.505, 0.515$, and $0.525$, respectively.
\end{itemize}
In this way, we create overlapping 2D projected bins of galaxies. Allowing these overlaps serves two purposes: 1) to better trace the evolution of clustering patterns across redshift, and 2) to reduce information loss caused by boundary effects between adjacent redshift bins.

\subsection{COLA Mock Catalogues}

COLA~\citep{Tassev+etal+2013} is a hybrid simulation method that integrates second-order Lagrangian perturbation theory (2LPT) and N-body algorithms, offering an effective solution for simulating dark matter (DM) particles. The perturbation theory results, which can successfully describe large-scale clustering processes, can replace the time integration calculations of N-body simulations on linear scales, so the large-scale and small-scale evolution processes are decoupled and separately treated. In the large-scale structure evolution, the evolution of DM particles can be described by the precise 2LPT theory results to describe their dynamical behaviors, while on small scales the evolution is carried out with well-established N-body simulation methods. To achieve this goal, COLA introduces a co-moving framework to place DM particles in a co-moving coordinate system that moves together with the ``LPT observer''. Although COLA sacrifices some accuracy on small scales, it can greatly improves computational efficiency by 10-100 times, while the accuracy on relatively large clustering scales is still quite good. 

In this study we used the mock SDSS CMASS galaxy catalogs produced in \cite{Ding+etal+2024}. The mocks are based on a set of simulation assumes a flat $\Lambda$CDM with $\Omega_m$ varying in region of $0.29 < \Omega_m < 0.33$, and the other parameter values fixed $\Omega_b=0.048$, $w=-1.0$, $\sigma_8=0.82$, $n_s=0.96$, and $h=0.67$, which are close to the average constraints of Planck 2015 results~\citep{Planck+etal+2016}. Each set of COLA simulation uses $1024^{3}$ particles in a cubic box with a side length of $800~h^{-1}$ Mpc. \cite{Ding+etal+2024} then applied SHAM method to COLA simulations to efficiently generate mock galaxies. Three parameters, including the scatter magnitude $\sigma_{\rm scat}$ in SHAM, the initial redshift of the COLA simulations $z_{\rm init}$ and the time step $da$, are tested to achieve a good match. A good agreement between COLA-simulated galaxies and CMASS NGC galaxies is achieved on scales ranging from 4 to 20 $h^{-1}{\rm Mpc}$ by setting $z_{\rm init} = 29$ and $da = 1/30$. Finally, by comparing various statistical properties such as anisotropic 2PCF, three-point clustering, and power spectrum multipoles, the authors found their COLA mocks achieve equally good performance when compared to GADGET mocks (in the region of $s\ge 4~h^{-1}{\rm Mpc}$ and $k\le 0.3~h{\rm Mpc}^{-1}$), while the computational cost is two orders of magnitude lower than the latter one. 

\begin{figure}
    \centering
    \includegraphics[width=0.68\textwidth]{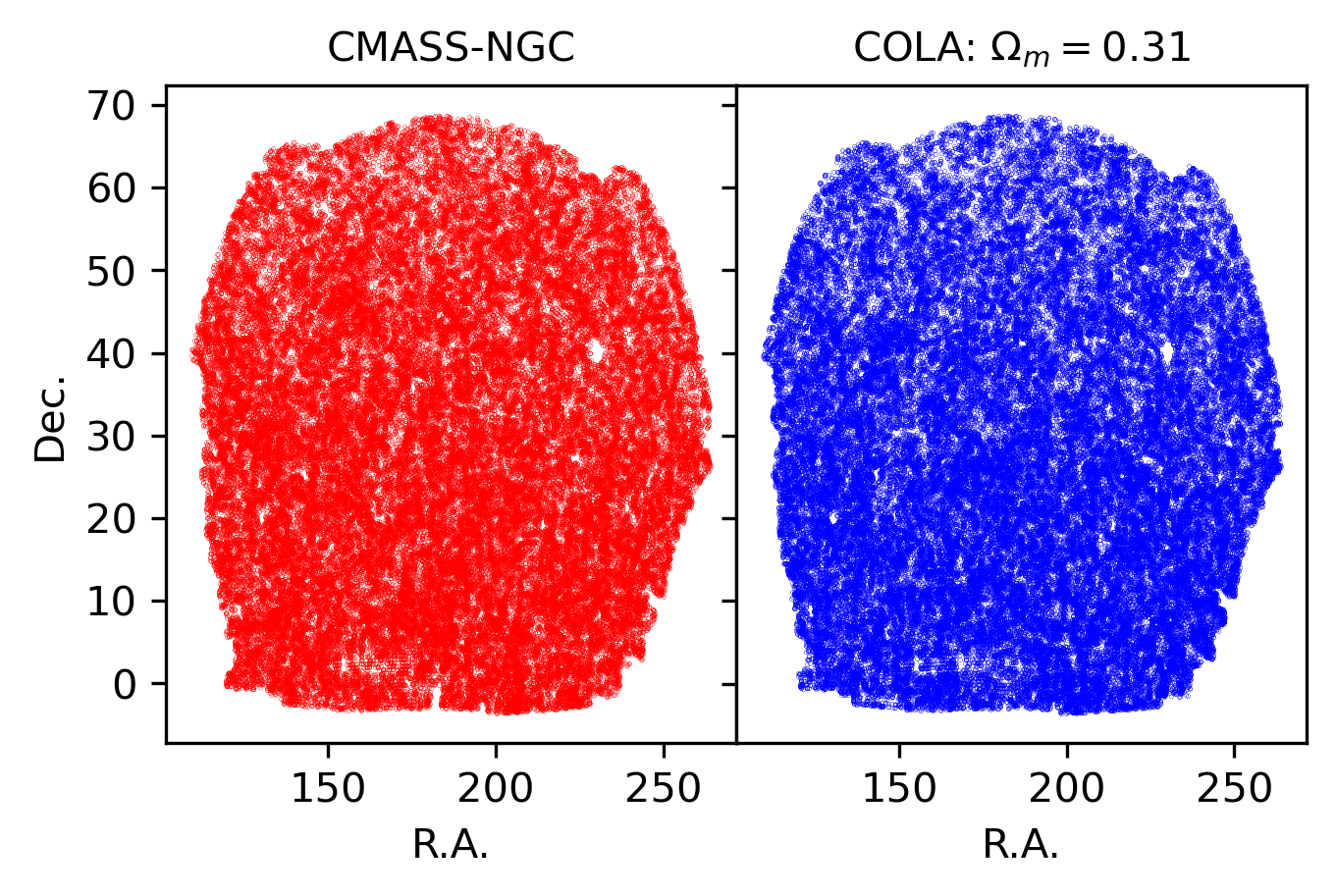}
    \caption{Comparison of the Right Ascension (R.A.)-Declination (Dec.) distribution of the CMASS-NGC galaxies (left panel) and galaxies in the $\Omega_m=0.31$ COLA mock (right panel). The 2D shell datasets are derived from the 3D samples within the redshift interval $z\in [0.48, 0.53]$, projected onto the central redshift plane $z_c=0.505$. The close agreement between the angular distributions of simulated and observed galaxies supports the validity of our simulation framework.}
    \label{fig:radec}
\end{figure}

Following the methodology of \cite{Yin+etal+2024}, in this proof-of-concept analysis we only used three sets of mock simulations with $\Omega_m=(0.25,0.31,0.4)$ to assess the constraining power of the statistics. Following the same procedure applied to observational data, we process each simulation by dividing them into six overlapping redshift bins and projecting onto 2D angular planes. For illustration purpose, Figure~\ref{fig:radec} compares the 2D positions of the observational and mock galaxies, in the $z\in [0.48, 0.53]$ subsample. This figure reveals very good similarity in angular galaxy distributions between COLA simulations and CMASS-NGC observations.

\subsection{PATCHY Mocks}

The PATCHY mock catalogs~\citep{Kitaura+etal+2016} combine efficient structure formation models with local, nonlinear, and scale-dependent stochastic biasing to generate realistic DM halo distributions. These mocks are constructed using augmented Lagrangian perturbation theory (2LPT for large scales plus spherical collapse for small scales), which evolves Gaussian density fluctuations into a full DM density field while self-consistently calculating peculiar velocities. The catalogs are carefully calibrated against the high-precision BigMultiDark N-body simulation~\citep{Kitaura+etal+2014,Rodr'iguez-Torres+etal+2016}, which contains $3840^3$ particles in a volume of $(2.5~h^{-1}{\rm Gpc})^3$ and adopts a $\Lambda$CDM cosmology with parameters $\Omega_m=0.307$, $\Omega_b=0.048$, $\sigma_8=0.82$, $n_s=0.96$, and $h=0.67$. The calibration is performed using analytic statistical biasing models.

For our analysis, we project the 3D PATCHY simulations into six distinct 2D shell layers that correspond to the CMASS-NGC observational slices. To reduce computational cost, we use only the first 300 realizations out of the total 2048 PATCHY mocks, which is sufficient for this proof-of-concept study. These mocks are then employed to compute the covariance matrices for the MACFs.

\section{Methodology}
\label{sect:analysis}

In this section, we present our methodology. We begin by outlining the theoretical framework and construction process of the $\beta$-skeletons, followed by an introduction to our novel mark-weighted correlation function statistics.

\subsection{Two-dimensional $\beta$-cosmic-web}

The $\beta$-skeleton algorithm originates from the fields of computational geometry and geometric graph theory. It is similar to the minimum spanning tree (MST) algorithm~\citep{Barrow+etal+1985}, with a key distinction being that the structural graph constructed relies on  $\beta$-parameter,  which makes it flexible in handling point sets with complex neighbor relationships~\citep{Kirkpatrick+Radke+1985, Correa+Lindstrom+2013}. 

\begin{figure}
    \centering
    \includegraphics[width=0.9\textwidth]{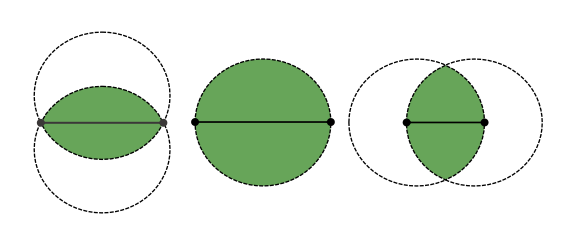}
    \caption{Lune-based definition of the empty regions of the $\beta$-skeleton. Left: $\beta<1$, Center: $\beta=1$, Right: $\beta>1$.}
    \label{fig:beta-theory}
\end{figure}

The definition of $\beta$-skeleton is as follows. For a point set $S$ in $n$-dimensional Euclidean space, any two points $p$ and $q$ in the point set $S$ are considered connected if there is no third point in the {\it empty regions}, which is illustrated in the Figure~\ref{fig:beta-theory}\footnote{The $\beta$-skeleton code  utilized in this study can be accessed at the following link: https://github.com/xiaodongli1986/LSSCode}:
\begin{itemize}
    \item For $0<\beta<1$, the empty region is formed by the intersection of two spheres with diameter $d_{pq} / \beta$, having $p$ and $q$ on their boundaries, where $d_{pq}$ is the common chord of the spheres. 
    \item For $\beta=1$, the empty region becomes a sphere with diameter $d_{pq}$. 
    \item For $\beta>1$, there are two different definitions of the empty region: the Circle-based definition and Lune-based definition. In this study we adopt the Lune-based definition, i.e. the empty region is the intersections of two spheres with a diameter of $\beta d_{pq}$ and their centers locating at $p+\beta(q-p) / 2$ and $q+\beta(p-q) / 2$, respectively.
\end{itemize}

The $\beta$-skeleton defined in this way exhibits a series of unique and interesting mathematical properties:
\begin{itemize}
    \item Hierarchical nature. As $\beta$ varies from 0 to $\infty$, the constructed $\beta$-skeleton graph transitions from a complete graph to an empty graph. In particular, the graphs formed by the special case of $\beta=1$ is called Gabriel Graph, which is known to contain the Euclidean minimum spanning tree, while in image analysis the  $\beta=1.7$ {\it Circle-based} graphs are found to accurately reconstruct the entire boundary of any smooth surface, without generating any edges that do not belong to the boundary, as long as the samples are sufficiently dense with respect to the local curvature of the surface.
    \item Inclusion relation. For any parameter value satisfying $\beta_1<\beta_2$, for fixed points $p$ and $q$, the empty region of the $\beta$-skeleton with $\beta_2$ contains the empty region of the $\beta$-skeleton with $\beta_1$ as the set parameter value. This property ensures that the connectivity of the web gradually decreases as increasing $\beta$. 
    \item Parametric control. The parameter $\beta$ provides an intuitive and powerful ``sparse-dense'' adjustment knob allowing researchers to finely tune the web's connection pattern according to the specific needs of scientific research, so that to conveniently visualize the web structures of the set. 
    \item Interpret-ability. $\beta$-skeleton is directly based on the geometric distribution of matter, which makes the results of the model highly intuitive and understandable. 
\end{itemize}

\begin{figure}
    \centering
    \includegraphics[width=0.9\textwidth]{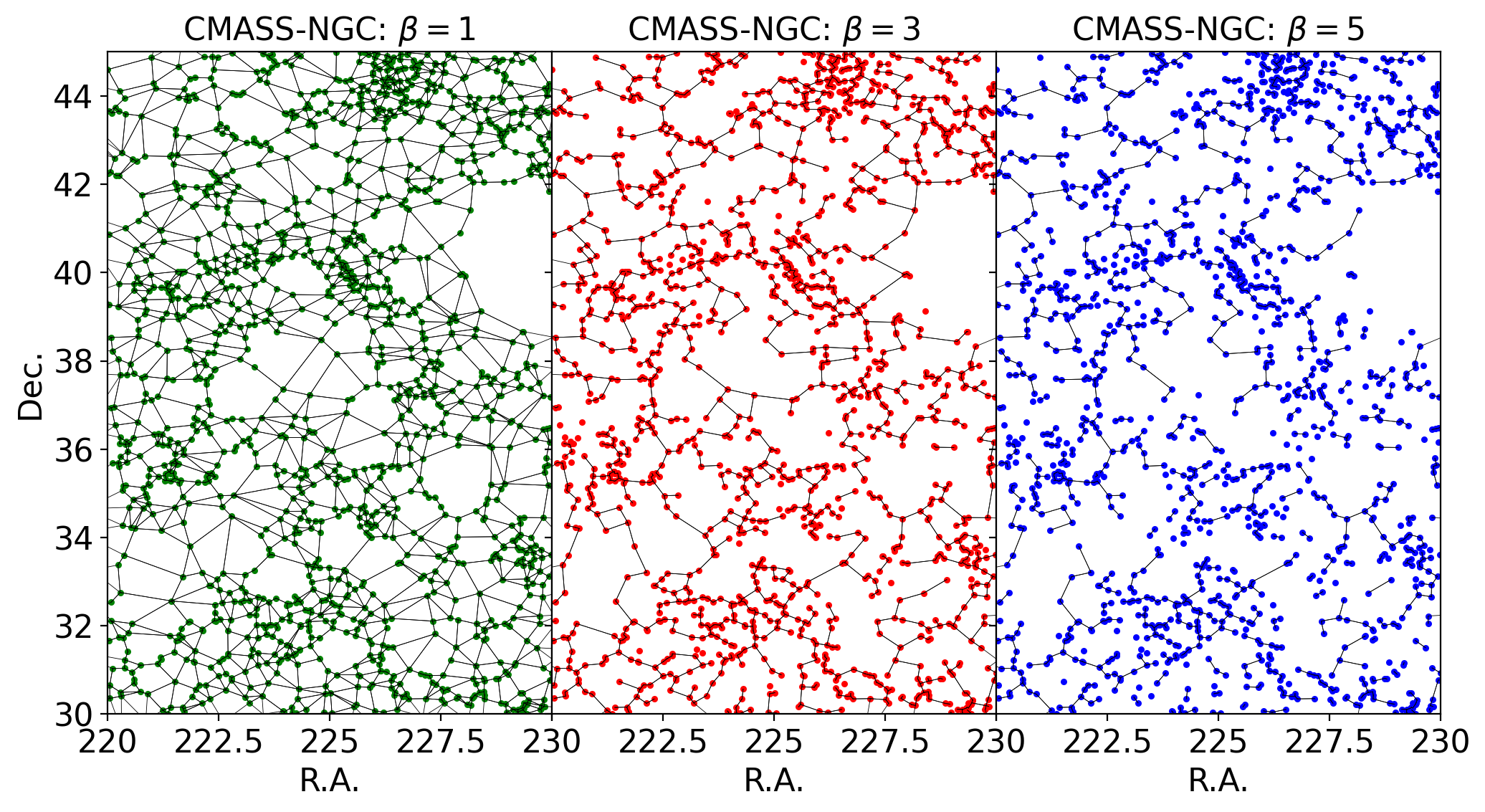}
    \caption{Visual comparisons of 2D $\beta$-cosmic-webs from CMASS-NGC data, with $\beta$ set to 1, 3 and 5, respectively. The structures are derived from galaxies with $z\in [0.48,0.53]$. Following~\cite{Yin+etal+2024}, we adopt the $\beta=3$ results as our baseline analysis, as its topological structure closely agrees with our conceptual understanding of the cosmic web.
    }
    \label{fig:CMASSweb}
\end{figure}

In Figure~\ref{fig:CMASSweb}, we visualize the 2D $\beta$-cosmic-web for CMASS-NGC galaxy data using $\beta = 1$, 3, and 5. The dataset includes galaxies within the redshift interval $[0.48, 0.53]$. For visualization purposes, we plot only 1500 galaxies with R.A. in $[220^\circ, 230^\circ]$ and Dec. in $[30^\circ, 45^\circ]$, corresponding to a slice of approximately $236.1\times 354.1~(h^{-1}\mathrm{Mpc})^2$. We observe that $\beta = 1$ produces a network with dense connections, whereas for $\beta = 5$, the connections become significantly sparser. Specifically, 
\begin{itemize}
    \item The $\beta=1$ web has 2695 links (overconnected). It contains a substantial number of extraneous and unnecessary galaxy connections that do not reflect true connection patterns.
    \item The $\beta=3$ web yields 1467 connections (nearly a 1:1 ratio with the number of galaxies) and is considered to realistically capture the connection relationships among galaxies.
    \item The $\beta=5$ web reduces to 1150 links (relatively sparse), omitting rich structural details.
\end{itemize}
Following \cite{Yin+etal+2024}, we adopt the $\beta=3$ web as the foundation for our analysis, as its topological structure aligns well with the commonly accepted conceptualization of the cosmic web.

\begin{figure}
    \centering
    \includegraphics[width=0.9\textwidth]{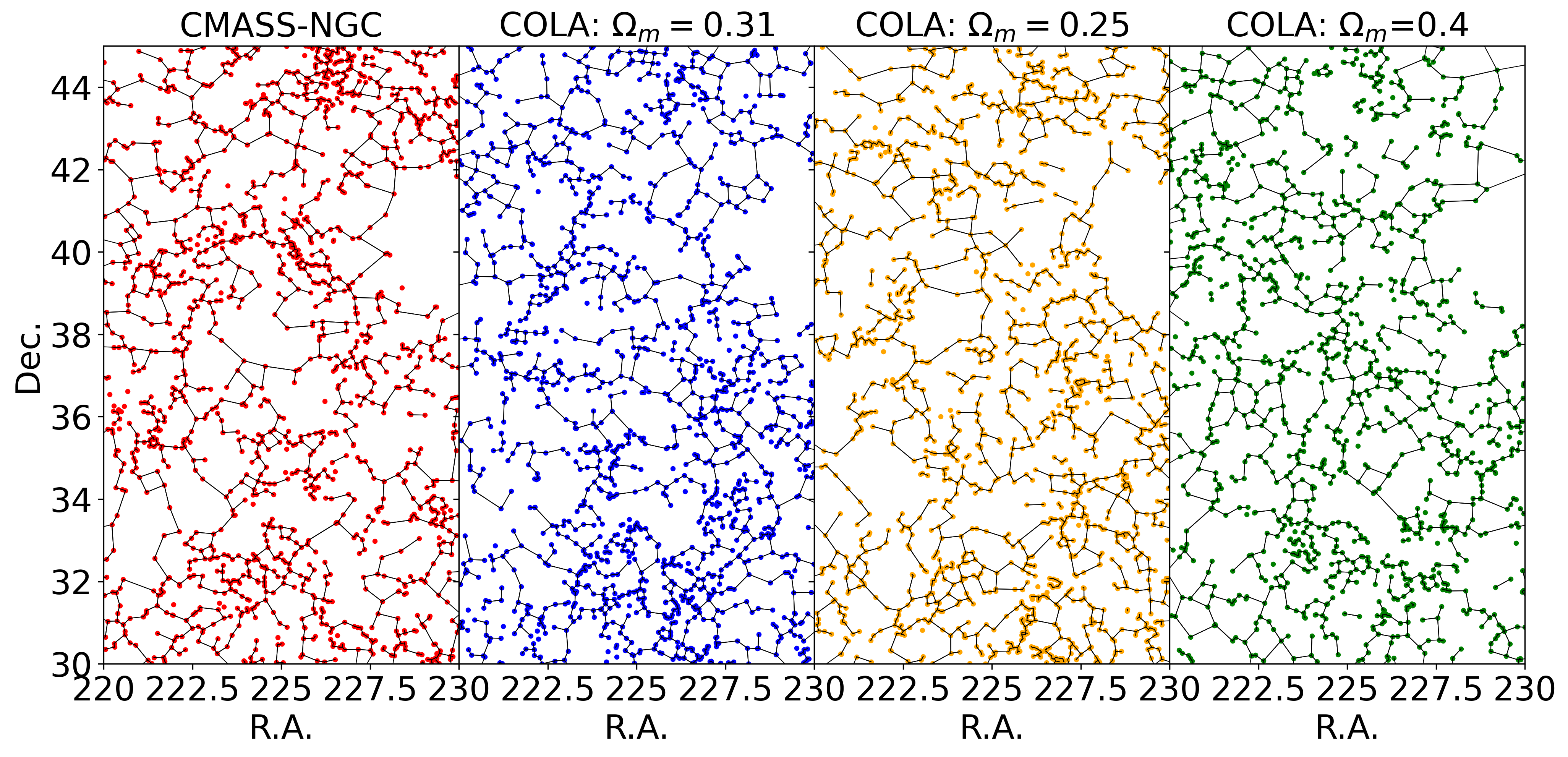}
    \caption{The $\beta=3$ web of the CMASS-NGC galaxies and the COLA mocks with $\Omega_m=0.31, 0.25, 0.4$. In all plots we chose galaxies with  $z\in [0.48,0.53]$, R.A. $\in [220^\circ, 230^\circ]$ and Dec. $\in [30^\circ, 45^\circ]$. By comparing the the connectivity of these webs, we find COLA simulation with $\Omega_m=0.31$ shows the best agreement with the observational data.}
    \label{fig:fourweb}
\end{figure}

Figure~\ref{fig:fourweb} displays the $\beta=3$ webs derived from CMASS-NGC data and COLA simulations with varying $\Omega_m$ values ($0.31$, $0.25$, and $0.4$). Consistent with Figure~\ref{fig:CMASSweb}, only galaxies within the redshift range $z \in [0.48, 0.53]$, R.A. $\in [220^\circ, 230^\circ]$, and Dec.$\in [30^\circ, 45^\circ]$ are shown. This figure shows that the results of the $\Omega_m=0.31$ mock is statistically most consistent with the observations. In terms of connectivity, the cosmic web constructed from the CMASS-NGC subsample exhibits 1467 connections. The mock catalog for $\Omega_m = 0.31$ yields a comparable number, with 1534 connections. In contrast, larger discrepancies are observed for the mocks with $\Omega_m = 0.25$ and 0.4, which produce 1548 and 1359 connections, respectively.

\begin{figure}
    \centering
	\includegraphics[width=0.9\textwidth]{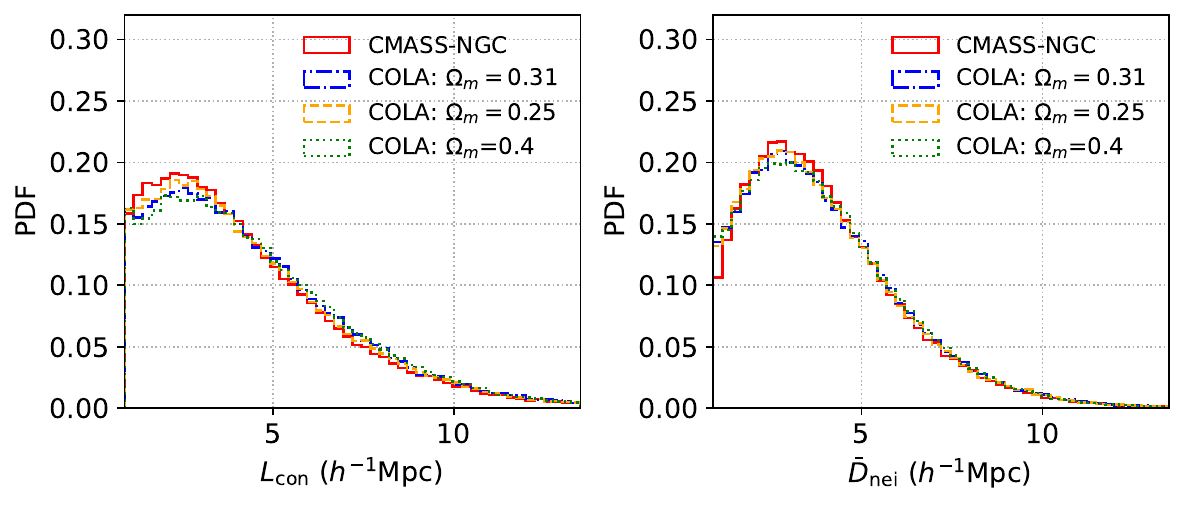}
    \caption{Comparison of the probability distributions of connection length, $L_{\rm{con}}$ (left), and the averaged connection length between each galaxies, $\bar D_{\rm{nei}}$ (right), in 2D $\beta$-cosmic-webs of $\beta=3$ for CMASS-NGC and COLA simulations with $\Omega_m=0.31$, 0.25, and 0.4.}
    \label{fig:statistics}
\end{figure}

We compare the $\beta$-cosmic-web statistics derived from COLA simulations and observational data in Figure~\ref{fig:statistics}. Although some differences are present, they are relatively minor--intuitively, the curves lie too close to each other to reliably distinguish between different cosmological models. Moreover, the shapes of the probability distributions of $L_{\rm con}$ and $\bar D_{\rm nei}$ are influenced by various systematics, making direct cosmological inference difficult~\citep{Fang+etal+2019,García+etal+2020}. As a result, the corresponding probability distributions are not well-suited for placing strong cosmological constraints. Instead, the most promising application of the $\beta$-cosmic-web in cosmological analysis remains the MCFs~\citep{Yin+etal+2024}.

\subsection{Mark weighted angular correlation functions using 2D $\beta$-cosmic-web}

In brief, the procedure for computing MCFs follows the standard 2PCF framework, with the key difference being the incorporation of weights derived from the $\beta$-skeleton statistics. While the traditional 2PCF is defined as  
\begin{equation}
\xi(\boldsymbol{r}) = \langle \delta(\boldsymbol{x}) \delta(\boldsymbol{x} + \boldsymbol{r}) \rangle,
\end{equation}  
the MCF takes the form  
\begin{equation}
W(\boldsymbol{r}) = \left\langle \delta(\boldsymbol{x})\, f(\boldsymbol{x})\, \delta(\boldsymbol{x} + \boldsymbol{r})\, f(\boldsymbol{x} + \boldsymbol{r}) \right\rangle,
\end{equation}  
where $f(\boldsymbol{x})$ represents the assigned weights, and $\delta(\boldsymbol{x}) = \delta \rho / \bar{\rho}$ denotes the number density contrast.

Following the methodologies developed in \cite{Yin+etal+2024}, we use three different quantities in the weighting:\begin{itemize}
    \item  Connection number ($N_{\rm con}$): the number of neighboring galaxies around each galaxy.
    \item  Mean neighbor distance ($\bar{D}_{\rm nei}$): the average distance between each galaxy and its neighboring galaxies.
    \item Inversed neighbor distance ($1/\bar{D}_{\rm nei}$): the reciprocal of mean neighbor distance.
\end{itemize}

Therefore, the weighting scheme is defined as
\begin{equation}
f(\boldsymbol{x})=\left\{\begin{array}{l}
N_{\rm con}(\boldsymbol{x})\,,  \\
\bar{D}_{\rm nei}(\boldsymbol{x})\,, \\
1 /\bar{D}_{\rm nei}(\boldsymbol{x})\,,
\end{array}\right.
\label{eq:weightrho}
\end{equation}
where different choices correspond to different physical interpretations: the number of connections $N_{\rm con}$, the average neighbor distance $\bar{D}_{\rm nei}$, or its reciprocal. Isolated galaxies (i.e., those without neighbors) are assigned a weight of $f(\boldsymbol{x}) = 0$. In particular, the reciprocal distance measure $1/\bar{D}_{\rm nei}$ in 2D projections can amplify extreme values and potentially lead to divergence. To ensure numerical stability, we introduce a cutoff threshold $f_{\rm cut}$, defined as the 95th percentile of the $1/\bar{D}_{\rm nei}$ distribution. Values exceeding this threshold are truncated at $f_{\rm cut}$, effectively suppressing outliers while preserving key nonlinear clustering features.

\begin{figure}
    \includegraphics[width=\textwidth]{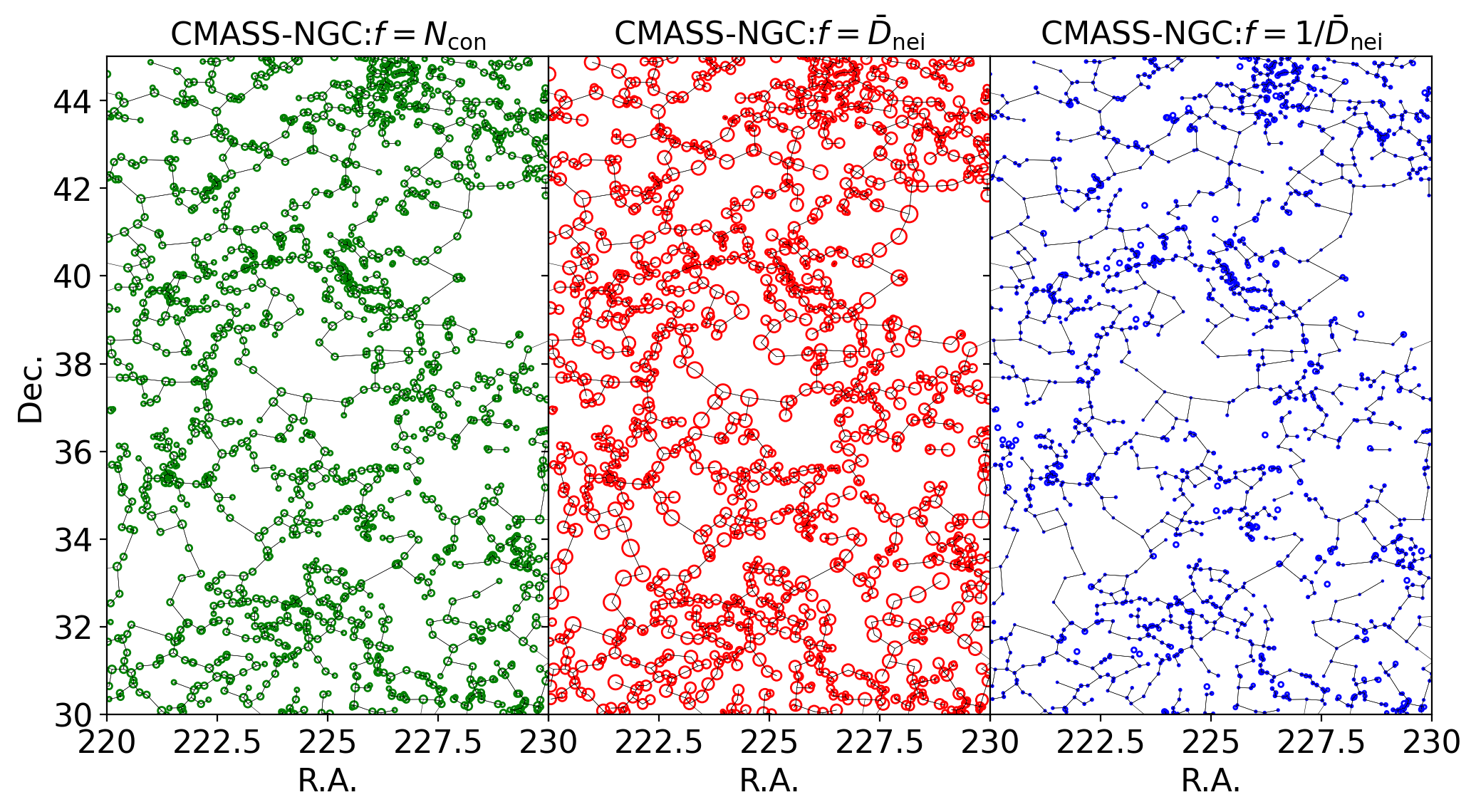}
    \caption{Visualization of the 2D $\beta$-cosmic-web weights for CMASS-NGC galaxies with $\beta=3$, represented by circle sizes. The region spans R.A. $\in [220^\circ, 230^\circ]$ and Dec. $\in [30^\circ, 45^\circ]$. The weights are derived from a 2D shell projected from 3D spatial data within redshift slice $\Delta z = 0.05$ at $z \in [0.48, 0.53]$, centered at $z_c = 0.505$. Three weighting schemes are shown side-by-side, from left to right: $N_{\rm con}$, $\bar{D}_{\rm nei}$, and $1/\bar{D}_{\rm nei}$, corresponding respectively to uniform weighting, emphasis on low-density environments, and strong prioritization of high-density structures. }
    \label{fig:wei}
\end{figure}

Figure~\ref{fig:wei} visualizes 2D $\beta$-cosmic-web weights of CMASS-NGC observational galaxies. The weights of galaxies are denoted by the circle sizes. In different weighting schemes, we find different environment-dependence of the weights. In the left panel, the sizes of the circles are not very different from each other, indicating that the dependence of $N_{\rm con}(\boldsymbol{x})$ on the environment is weak.  When $\bar{D}_{\rm nei}(\boldsymbol{x})$ is used as weight, it upweights the low-density regions, while its inverse $1/\bar{D}_{\rm nei}(\boldsymbol{x})$ highlights the overdense regions.

Based on these weights, MCFs are then computed through Landy–Szalay estimator~\citep{Landy+etal+1993}:
\begin{eqnarray}
    W(s,\mu) = \frac{WW-2WR+RR}{RR}\,,
\end{eqnarray}
where $WW$ denotes the weighted count of galaxy-galaxy pairs, $WR$ corresponds to galaxy-random pairs, and $RR$ represents random-random pairs. $s$ represents the distance between pairs, and $\mu$ is defined as cosine of the angle between the line of sight (LOS) direction and the line connecting the pair galaxies.

The monopole of MCF, which depends only on the clustering scale $s$, is obtained by integrating $W(s, \mu)$ over all angles:
\begin{equation}\label{eq:ws}
W_0(s) = \int_0^1 W(s, \mu) \, d\mu.
\end{equation}

Assuming a flat cosmological model, the comoving distance between galaxies at redshifts $z_1$ and $z_2$ respectively can be expressed as
\begin{equation} s=|\bm{s}|=\sqrt{r^2(z_1)+r^2(z_2)-2r(z_1)r(z_2)\cos{\theta}}\,,
\end{equation}   
where $r(z_1)$, $r(z_2)$ represent the comoving distances from the observer for two galaxies at redshifts of $z_1$ and $z_2$, respectively, and $\theta$ is the angular distance between galaxy pairs. In $\Lambda$CDM model, the comoving distance $r(z)$ of a galaxy with redshift takes form of,
\begin{equation}
    r(z)=\frac{1}{H_0}\int_0^z\frac{dz'}{\sqrt{\Omega_{\Lambda}+\Omega_m(1+z')^3}}\,,
    \label{equ:r1}
\end{equation}
where $H_0$ is the present-day Hubble constant, and $\Omega_{\Lambda}$ is the dark energy density parameter.

Under the flat-sky approximation and thin redshift slice assumption, for a set of galaxies with the center of redshift bin denoted as $z_{c}$, the angular component of the comoving distance between galaxy pairs take the form of
\begin{equation}\label{eq:sp}
   s\approx  \sqrt{2r^2(z_c)-2r^2(z_c)\cos{\theta}}=2r(z_c)\sin{(\frac{\theta}{2})}\approx r(z_c)\theta\,,
\end{equation} 
where $\theta$ represents the angular separation between the two galaxies. The last approximation is valid when $\theta \ll 1$, which holds in our analysis.

Since we consider a fixed redshift interval of $\Delta z = 0.05$ and angular separations across 6 redshift bins, all within the range $\theta \in [0.43^\circ, 2.62^\circ]$,  the flat-sky approximation and thin redshift slice assumptions are well satisfied. Consequently, the clustering scale $s$ is directly related to the angular separation $\theta$ between two galaxies, as given by Equation~\ref{eq:sp}.

In this study, we focus exclusively on the monopole component, $W_0$. The 2D $\beta$-cosmic-web-based MACFs, denoted by $\hat{w}(\theta)$, are computed using a normalized form (to mitigate the influence of galaxy bias and improve the accuracy of the analysis) defined as
\begin{equation}\label{eq:wsmu}
    \hat{w}(\theta) = \frac{W_0(s)}{\int_a^b W_0(s)\,ds},
\end{equation}
where $\theta = s / r(z_c)$, in accordance with Equation~\ref{eq:sp}.  We examined the impact of varying the integration limits $a$ and $b$, and adopted $a = 10~h^{-1}\mathrm{Mpc}$ and $b = 58~h^{-1}\mathrm{Mpc}$ as our fiducial values. This range reflects a conservative compromise, chosen to retain meaningful clustering information while mitigating potential systematic uncertainties. For the analysis, the clustering separation $s$ is divided into eight bins of equal width.

\begin{figure}
    \includegraphics[width=\textwidth]{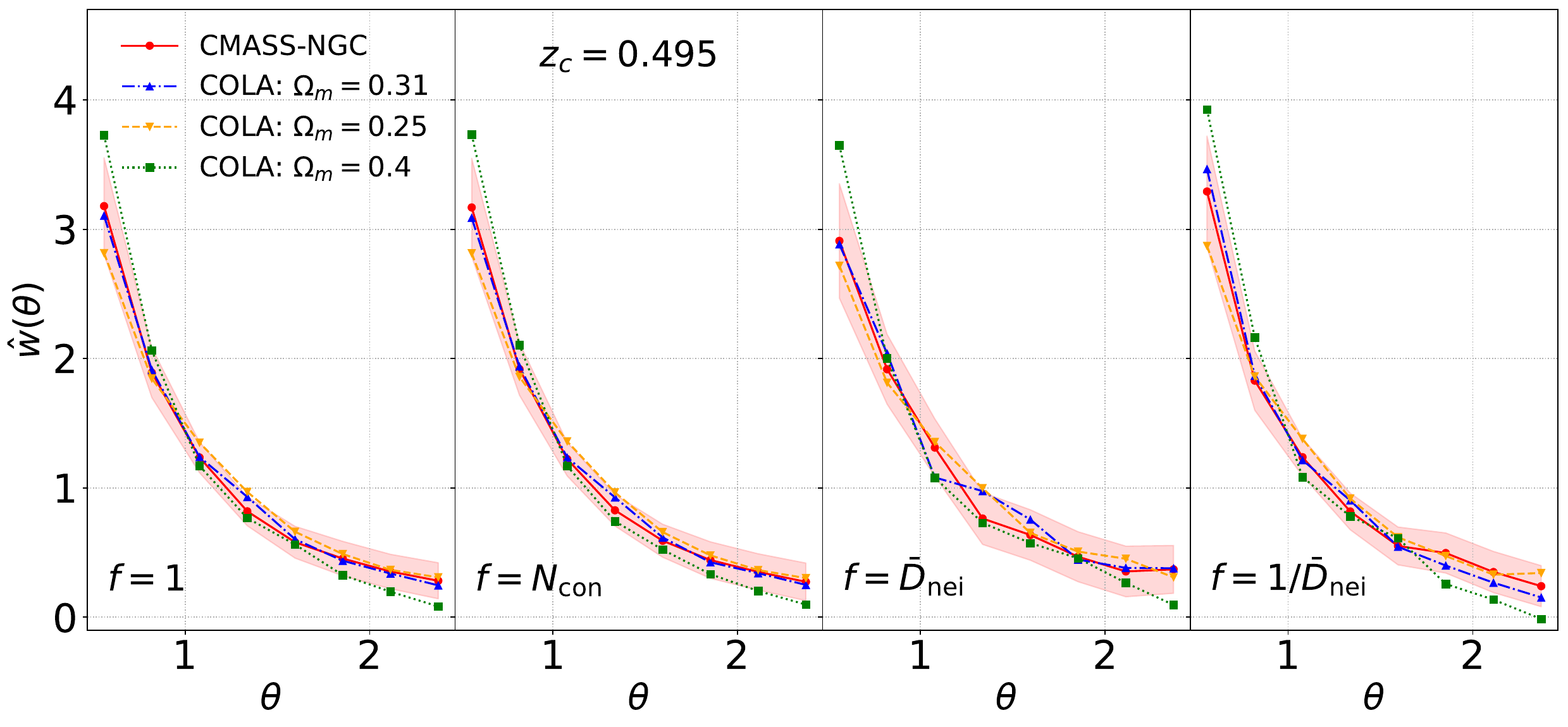}
    \caption{Comparison of normalized angular correlation functions $\hat w(\theta)$ for CMASS-NGC (red) and COLA simulations with $\Omega_m=0.31$ (blue), $0.25$ (orange), $0.4$ (green) using galaxies with $z\in[0.47,0.52]$, computed in the range of $\theta \in [0.45^\circ,2.52^\circ]$. The shaded region represents 3$\sigma$ confidence interval. Results of the $\Omega_m=0.31$ COLA mock is most consistent with the observational data, while both $\Omega_m=0.25$ or $\Omega_m=0.4$ shows significant discrepancies ($>3\sigma$) from the deviate from the CMASS-NGC data.
    }
    \label{fig:MCFz0.495}
\end{figure}

\begin{figure}
    \includegraphics[width=\textwidth]{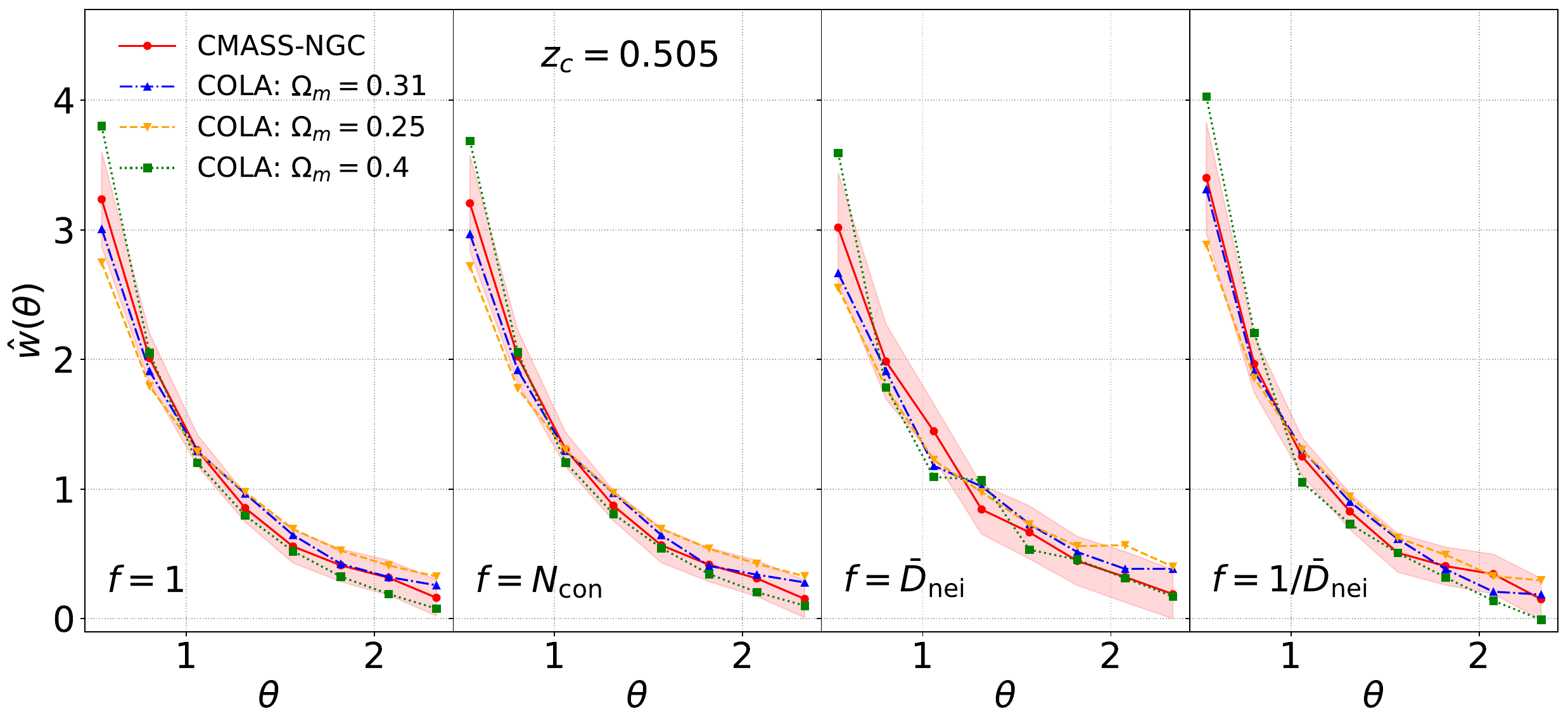}
    \caption{Same as in Figure~\ref{fig:MCFz0.495}, but for measurements in the redshift interval $[0.48, 0.53]$. We still find that the mock with $\Omega_m = 0.31$ best agrees with the observational data.
}
    \label{fig:MCFz0.505}
\end{figure}

Figure~\ref{fig:MCFz0.495} compares  the $\hat{w}(\theta)$ values measured from CMASS-NGC galaxies (red) and from COLA mocks with $\Omega_m = 0.31$ (blue), $0.25$ (orange), and $0.4$ (green), using galaxies in the redshift range $z \in [0.47, 0.52]$. Measurements from the standard 2PACF and the MACFs weighted by $N_{\rm con}$, $\bar{D}_{\rm nei}$, and $1/\bar{D}_{\rm nei}$ are shown. Note that, the case $f= 1$ is equivalent to 2PACF measurements. COLA with $\Omega_m=0.31$ aligns with CMASS-NGC within the $3\sigma$ confidence interval, while $\Omega_m=0.25$ or $\Omega_m=0.4$ simulations show evident discrepancy from the data. The trend is similar for other redshift intervals (e.g. Fig.~\ref{fig:MCFz0.505} compares the measurements of galaxies with $z\in [0.48,0.53]$).

To assess the constraining power of different statistics, we define the $\chi^2$ as 
\begin{equation}\label{eq:chi1}
    \chi^2=(\Delta\boldsymbol{p})^T\cdot \boldsymbol{C}^{-1}\cdot\Delta\boldsymbol{p}\,,
\end{equation}
where $ \Delta\boldsymbol{p}$ characterizes the difference between the measurements from the data ($\boldsymbol{p}_{\rm data}$) and the COLA mocks ($\boldsymbol{p}_{\rm model}$)
\begin{equation}
    \Delta\boldsymbol{p}=\boldsymbol{p}_{\rm model}-\boldsymbol{p}_{\rm data}\,.
\end{equation}
The covariance matrix corresponding to the normalized weighted angular correlation function statistic $\hat w(\theta)$ is represented by $\boldsymbol{C}$. By varying the value of $\Omega_m$, the variation of the $\chi^2$ values can be utilized to evaluate the sensitivity of a specific statistics to the cosmological parameter, thereby enabling us to compare the power of different statistics. The covariance matrix is computed through
\begin{eqnarray}
\boldsymbol{C} &=& \left<\left(\boldsymbol{p}-\boldsymbol{\bar{p}}\right)\left(\boldsymbol{p}-\boldsymbol{\bar{p}}\right)^T \right> \\
&=&
\frac{1}{N_{\rm mock}-1} \sum_{i=1}^{N_{\rm mock}}  \left(\boldsymbol{p}_i - \boldsymbol{\bar{p}}\right) \left(\boldsymbol{p}_i-\boldsymbol{\bar{p}}\right)^T\,.
\end{eqnarray}
Here, the vector $\boldsymbol{p}_i$ denotes the $i$-th PATCHY mock out of a total of $N_{\rm mock}$ mocks, and $\boldsymbol{\bar{p}}$ represents the average of $\boldsymbol{p}$ over all PATCHY mocks. In this analysis, we employ 300 PATCHY simulation sets to estimate the covariance matrix. The vector $\boldsymbol{p}$ collects all measured quantities. For a given $z$-bin, each statistic yields 8 measured quantities, so combining two weighting schemes results in a total of 16 measured elements. Therefore, the number of PATCHY simulations is sufficient to accurately estimate the covariance matrix.

To reduce statistical fluctuations and enhance the robustness of our results, we use the average of the $\chi^2$ values from six subsamples (each covering different but overlapping redshift ranges) to assess the performance of different statistics:
\begin{equation}\label{eq:chi2}
    \overline{\chi^2}=\frac{1}{n}\sum_{i=1}
    ^n \chi_i^2\,,
\end{equation}
where $n = 6$, and $\chi_i^2$ denotes the chi-square value calculated for the $i$-th individual 2D shell of galaxies.

\section{Results}

\begin{figure}
    \centering
	\includegraphics[width=0.85\textwidth]{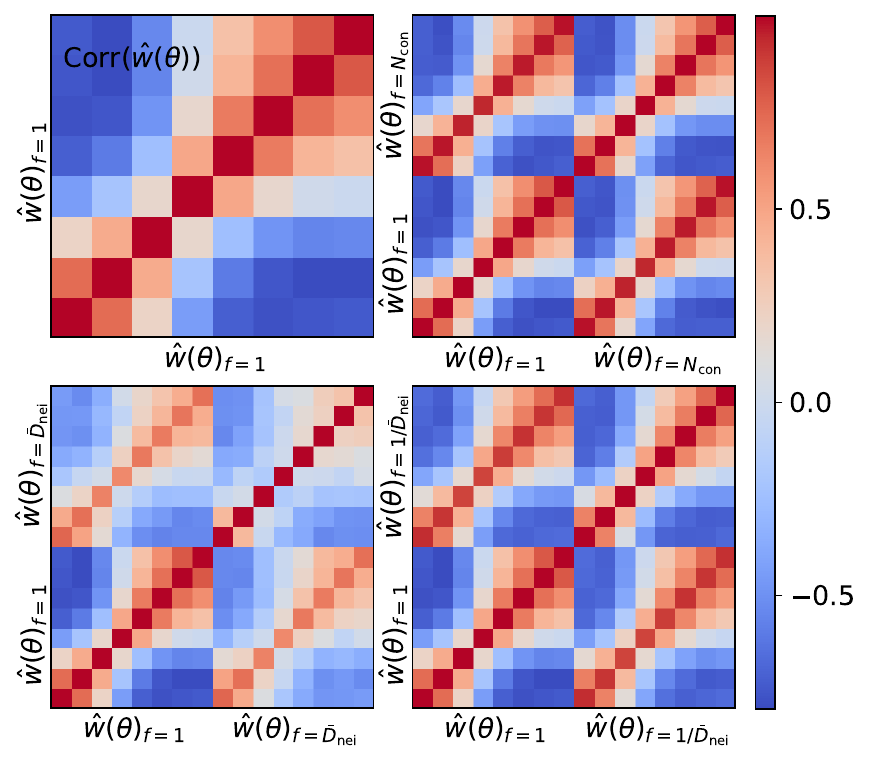}
    \caption{Normalized covariance matrix showing the correlation coefficients of $\hat{w}(\theta)$ computed for various combinations with different weighting schemes, obtained from PATCHY simulations.
 Here, we show results only for the sample with $z \in [0.48, 0.53]$.
}
    \label{fig:corr}
\end{figure}

Figure~\ref{fig:corr} shows the correlation coefficient of the $z\in [0.48,0.53]$ subsample. To precisely quantify the relationships between different statistical measures, the figure displays both the autocorrelation of the correlation functions and the cross-correlation coefficients between the standard 2PACF and the $\hat w(\theta)$ weighted by $N_{\rm con}$, $\bar{D}_{\rm nei}$, and $1/\bar{D}_{\rm nei}$, respectively. Among them, the MACF weighted by $N_{\rm con}$ is most correlated with 2PACF, since this weighting scheme is most close to uniform weighting.

The sensitivity of $\hat w(\theta)$ to cosmological parameter is then evaluated using
\begin{equation}\label{eq:s/n}
\Delta \chi^2\equiv\chi^2 -\chi^2_{\rm min}\,,
\end{equation}
which directly measures the statistical power to discriminate between different cosmologies. To enhance the stability and robustness of our results, we compute the mean $\Delta \chi^2$ cross the six overlapping redshift bins, defined as:
\begin{equation}\label{eq:s/n1}
\Delta \overline{\chi^2}\equiv\overline{\chi^2 }-\overline{\chi^2_{\rm min}}\,,
\end{equation}
where $\overline{\chi^2}$ and $\overline{\chi^2_{\rm min}}$ represent the average $\chi^2$ values over all bins, as defined in Equation~\ref{eq:chi2}.
This averaging procedure serves two purposes: 1) to reduce statistical fluctuations arising from the relatively small sample size, and 2) to improve robustness against random variations in our proof-of-concept analysis, which currently includes only three simulations. By adopting this approach, we effectively suppress statistical noise while preserving sensitivity to cosmological parameters.

\begin{figure}
    \centering
	\includegraphics[width=0.5\textwidth]{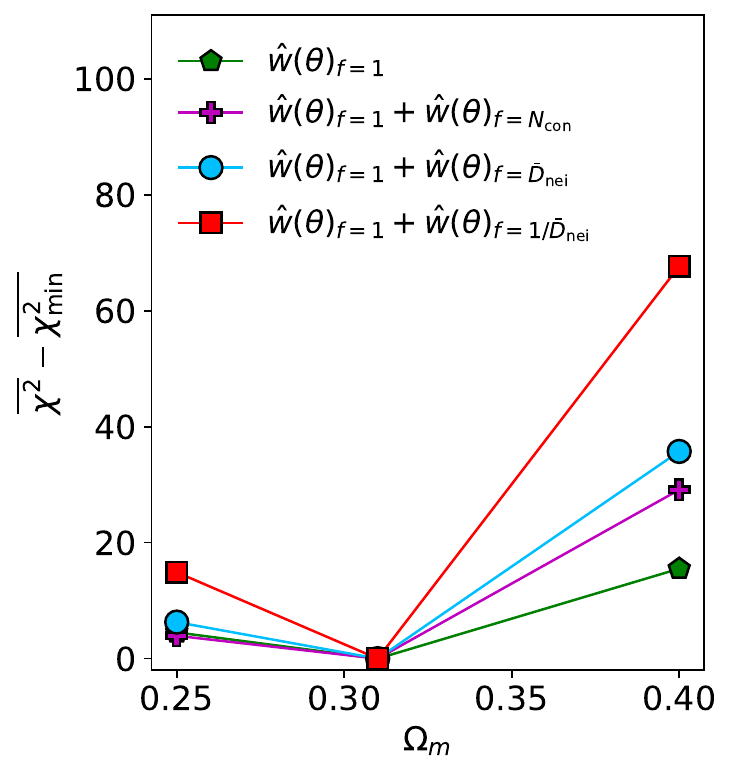}
    \caption{The variation of $\Delta\overline{\chi^2}$ as $\Omega_m$ changes from 0.25 to 0.4. It reveals a minimum $\overline{\chi^2_{\rm min}}$ at $\Omega_m=0.31$, distinctively lower than values at other two $\Omega_m$.  Combination of 2PACF and $\hat w(\theta)$ weighted by $1/\bar{D}_{\rm nei}$ (represented by red squares) yields to the highest sensitivity to $\Omega_m$.}
    \label{fig:chi}
\end{figure}

Figure~\ref{fig:chi} presents the variation of $\Delta \overline{\chi^2}$ as $\Omega_m$ varies from 0.25 to 0.4. In all cases we find $\Omega_m=0.31$ leads to the minimal $\chi^2$, so it always serves as the baseline. We find that:
\begin{itemize}
    \item the incorporation of weighted statistics provides significant improvement over using the two-point angular correlation function (2PACF) alone, as evidenced by the enhanced $\Delta \overline{\chi^2}$. This demonstrates greater statistical power for constraining cosmological parameters.
    \item among all tested combinations, the most sensitive probe of $\Omega_m$ is achieved by combining 2PACF with the $\hat{w}(\theta)$ statistic weighted by $1/\bar{D}_{\rm nei}$ (red squares in Fig.~\ref{fig:chi}), enlarging the value of $\Delta \chi^2$ by 2-3 times. 
\end{itemize}
To quantitatively assess the improvement gained by incorporating $\beta$-cosmic-web-weighted $\hat{w}(\theta)$ statistics compared to the conventional 2PACF approach, we introduce the relative sensitivity metric:
\begin{equation}\label{eq:rs}
r= \frac{\Delta\overline{\chi^2}}{\Delta \overline{\chi^2}_{\rm 2PACF}}-1\,,
\end{equation}
where $\Delta \overline{\chi^2}_{\rm 2PACF}$ refers to the results obtained using only the 2PACF measurements, that is, by calculating $\hat{w}(\theta)$ using the weighting scheme $f = 1$ as defined in Equation~\ref{eq:wsmu}.

The statistical improvements are summarized in Table~\ref{tab:chi11}. Our analysis reveals that combining 2PACF with MACFs significantly enhances the statistical confidence. Specifically:
\begin{itemize}
\item Incorporating $\bar D_{\rm nei}$-weighted statistics boosts the $\Delta \overline{\chi^2}$ values by 39\%–130\%, indicating substantially improved parameter constraints.
\item The $1/\bar D_{\rm nei}$ weighting scheme shows even more dramatic improvement, enhancing $\Delta  \overline{\chi^2}$ by 229\%–336\%. Statistically, this enhancement is equivalent to increasing our dataset by the same amount.
\end{itemize}

\begin{table}[ht]
\centering
\begin{tabular}{lll}
\hline
\hline
Statistic   & $\Delta\overline{\chi^2}$ / $r$ ($\Omega_m = 0.25$) & $\Delta\overline{\chi^2}$ / $r$ ($\Omega_m = 0.4$) \\ 
\hline
$\hat w(\theta)_{f=1}$                           & 4.52 / 0\%    & 15.51 / 0\%      \\  
$\hat w(\theta)_{f=1}+\hat w(\theta)_{f=N_{\rm con}}$        & 3.94 / –13\%  & 29.12 / +88\%    \\
$\hat w(\theta)_{f=1}+\hat w(\theta)_{f=\bar{D}_{\rm nei}}$  & 6.30 / +39\%  & 35.73 / +130\% \\
$\hat w(\theta)_{f=1}+\hat w(\theta)_{f=1/\bar{D}_{\rm nei}}$  & 14.89 / +229\%  & 67.68 / +336\% \\ 
\hline
\hline\\
\end{tabular}
\caption{
Sensitivity of each statistic to $\Omega_m$ is quantified by $\Delta \overline{\chi^2}$ and the relative deviation $r$ (Eq.~\ref{eq:rs}). The percentage values indicate the improvement in $\Delta \overline{\chi^2}$ relative to the  2PACF results, indicated by $\hat w(\theta)_{f=1}$. The $1/\bar{D}_{\rm nei}$-weighted $\hat{w}(\theta)$ yields the largest enhancement, with improvements of 229\% at $\Omega_m = 0.25$ and 336\% at $\Omega_m = 0.4$.
}
\label{tab:chi11}
\end{table}

\section{Concluding remarks}\label{sect:conclude}

In this study, we investigate whether cosmological constraints can be improved using the $\beta$-cosmic-web weighted angular correlation functions. We compare the statistical measurements from the SDSS DR12 CMASS-NGC galaxies and the COLA mocks with different values of $\Omega_m$ (0.25, 0.31 and 0.4), and use different statistics to distinguish between these $\Lambda$CDM cosmologies. We quantify the statistical power of each method using the value of $\Delta \overline{\chi^2}$.

Our results show that adding the $\beta$-cosmic-web weighted statistics can effectively enhance the statistical power. Specifically, weighting the statistics by the mean neighbor distance ($\bar D_{\rm nei}$) increases the $\Delta \overline{\chi^2}$ values by approximately 40\%–130\%. An even more significant enhancement is achieved with the inverse mean neighbor distance weighting ($1/\bar D_{\rm nei}$), which boosts $\Delta \overline{\chi^2}$ by a factor of 2–3 (corresponding to an improvement of about 230\%–340\%) compared to unweighted traditional angular statistics.

The procedure of this analysis follows that of \citet{Yin+etal+2024}, who investigated the statistical power of $\beta$-cosmic-web weighted correlation functions. Their results showed that including a single weighted statistic enhances $\Delta \overline{\chi^2}$ by factors of a few, with multiple weights yielding improvements approaching an order of magnitude.  Consistent with these findings, the present study also demonstrates that the $1/\bar{D}_{\rm nei}$ weighting scheme performs best, indicating that dense regions carry substantial cosmological information. In contrast, weighting by $N_{\rm con}$ provides minimal improvement. Although this study focuses on constraints of $\Omega_m$, it is expected that these statistics are sensitive to other cosmological parameters, such as those related to dark energy~\citep{Riess+etal+1998, Perlmutter+etal+1999, Weinberg+1989,Li+etal+2011,Yoo+Watanabe+2012,Weinberg+etal+2013, Wang+Gu+etal+2024, Gu+Wang+etal+2024}.

The statistics developed in this study are based on thin redshift slices, making them well-suited for slitless spectroscopic surveys (e.g., Euclid, CSST) and wide-band photometric surveys, where redshift uncertainties hinder traditional 3D clustering. The 2D angular correlation function offers greater robustness to such uncertainties, enabling broader applicability. Additionally, since marked statistics have not been previously explored using 2D angular clustering, this study also provides new theoretical insights.

In this proof-of-concept study, we only considered three weighting schemes. Since the dependence of clustering on the environment is complex and varied, we expect cosmological constraints could be further enhanced by exploring additional weighting approaches. For instance, the weighting scheme $\rho^\alpha$, which assigns weights based on the local density raised to the power $\alpha$, emphasizes either dense or underdense regions, improving the contrast between galaxy clusters in different environments and yielding tighter constraints~\citep{Yang+etal+2020,Lai+etal+2024}. Alternatively, weights based on the density gradient, $(|\nabla \rho|/\rho)^\alpha$, or combinations of both schemes can be used in 2D marked angular statistics~\citep{Xiao+etal+2022}. Additionally, cosmological constraints can be improved by employing reconstructed density fields~\citep{Wang+Shi+etal+2024, Shi+Wang+Yang+etal+2025}. To further advance slitless survey analyses, machine learning methods and scattering transforms applied to 2D redshift-sliced data also deserve exploration. These avenues will be pursued in future study.

\section{Acknowledgments}

This work is supported by the Ministry of Science and Technology of China (2020SKA0110401, 2020SKA0110402, 2020SKA0110100), the National Key Research and Development Program of China (2018YFA0404504, 2018YFA0404601, 2020YFC2201600), the National Natural Science Foundation of China (12373005, 11890691, 12205388, 12220101003, 12473097), the China Manned Space Project with numbers CMS-CSST-2021 (A02, A03, B01), Guangdong Basic and Applied Basic Research Foundation (2024A1515012309). This study utilized the Tianhe-2 supercomputer, the Kunlun cluster at Sun Yat-Sen University, and HPC resources from the Beijing Super Cloud Center (BSCC) and Beijing Beilong Super Cloud Computing Co., Ltd (http://www.blsc.cn/), all of which greatly contributed to the research.

\bibliographystyle{raa}
\bibliography{bibtex}

\end{document}